\documentclass[12pt,preprint]{aastex}

\usepackage{subfigure}

\begin{document}

\shorttitle{Spicules type~{\sc ii} simulations}
\shortauthors{Mart\'inez-Sykora \& Hansteen \& Moreno-Insertis}

\title{On the origin of the Type~{\sc ii} spicules - dynamic 3D MHD simulations}

\author{Juan Mart\'inez-Sykora $^{1,2}$}
\email{j.m.sykora@astro.uio.no}
\author{Viggo Hansteen $^{2}$} 
\email{viggo.hansteen@astro.uio.no}
\and
\author{Fernando Moreno-Insertis $^{3}$}
\email{fmi@iac.es}

\affil{$^1$ Lockheed Martin Solar and Astrophysics Laboratory, Palo Alto, CA 94304, USA}
\affil{$^2$ Institute of Theoretical Astrophysics, University of Oslo, P.O. Box 1029 Blindern, N-0315 Oslo, Norway}
\affil{$^3$ Instituto de Astrof\'isica de Canarias, 38200 La Laguna (Tenerife), Spain}

\newcommand{\myemail}{juanms@astro.uio.no}
\newcommand{\viscous}{\underline{\underline{\tau}}}
\newcommand{\resistive}{\underline{\underline{\eta}}}
\newcommand{\komment}[1]{\texttt{#1}}

\begin{abstract}

Recent high temporal and spatial resolution observations of the chromosphere
have forced the definition of a new type of spicule, ``type~{\sc ii}'s", 
that are characterized by rising rapidly, having short lives, and by fading 
away at the end of their lifetimes. Here, we report on features found 
in realistic 3D simulations of the outer solar atmosphere that resemble the
observed type~{\sc ii} spicules. These features evolve naturally from the 
simulations as a consequence of the 
magnetohydrodynamical evolution of the model atmosphere.
The simulations span from the upper layer of the convection 
zone to the lower corona and include the emergence of horizontal magnetic 
flux. The state-of-art {\em Oslo Staggered Code} (OSC) is used to solve 
the full MHD equations with non-grey and non-LTE radiative transfer and 
thermal conduction along the magnetic field lines. We describe in detail 
the physics involved in a process which we consider a possible candidate as a driver 
mechanism to produce type~{\sc ii} spicules. The modeled 
spicule is composed of material rapidly ejected from the chromosphere that 
rises into the corona while being heated. Its source lies in a region with
large field gradients and intense electric currents, which lead to a strong Lorentz force that
squeezes the chromospheric material, resulting in a vertical pressure gradient that
propels the spicule along the magnetic field, as well as Joule heating, which 
heats the the jet material, forcing it to fade.

\end{abstract}


\section{Introduction}

The upper solar chromosphere is filled with highly dynamic jets. 
These jet phenomena go under several names, depending on the location they are 
observed at and on their physical characteristics. 
Examples are dynamic fibrils found in active regions on the disk,
mottles found in the quiet sun on the disk, and spicules at the limb 
\citep{Beckers:1968qe,Suematsu:1995lr,Hansteen+DePontieu2006,de-Pontieu:2007hb,luc2007}.
Spicules at the limb are observed in the H$\alpha$ line and in other 
chromospheric lines such as the Ca {\sc ii} H \& K lines.
With the large improvement in spatio-temporal stability and resolution 
given by the Hinode satellite \citep{2007SoPh..243....3K},
and with the Swedish 1-m Solar Telescope (SST) \citep{Scharmer:2008zv}, it has
been discovered that spicules come in at least two clearly identified physical types
with different dynamics and timescales \citep{de-Pontieu:2007kl}.

The so-called type~{\sc i} 
spicules occur on time-scales of 3-10 minutes, develop speeds of 
10-30~km~s$^{-1}$, reach heights of 2-9~Mm \citep{Beckers:1968qe,Suematsu:1995lr},
and typically involve upward motion followed by downward motion. 
\cite{Shibata:1982fk,Shibata:1982qy} studied in detail the propagating 
shocks in simplified 1D models where they increased the temperature at 
specific heights and the velocities are not larger than the sound speed.
The position of the top of the spicule follows  a parabolic profile in time, which is 
indicative of an upwardly propagating shock passing through the upper 
chromosphere and transition region towards the corona as modeled and described 
by \citet{Hansteen+DePontieu2006}, \citet{De-Pontieu:2007cr}, 
\citet{Heggland:2007jt}, and in 3D by \citet{Martinez-Sykora:2009kl}. 
The latter authors described how the spicule-driving shocks can be generated 
by a variety of processes, such as collapsing granules, overshooting p-modes, 
dissipation of magnetic energy in the photosphere and lower-chromosphere, 
or any other sufficiently energetic phenomenon in the upper photosphere or chromosphere.  
\citet{Matsumoto:2010lr} claim that spicules can be 
driven by resonant Alfv\'en waves, generated in the photosphere and 
confined in a cavity between the photosphere and transition region.

Type~{\sc ii} spicules, observed in Ca~{\sc ii} and H$\alpha$, have shorter
lifetimes than type~{\sc i} spicules, typically less than 100 seconds, are
more violent, with upward velocities of the order of 50-100 km s$^{-1}$, and
reach somewhat greater heights.  They usually exhibit only 
upward motion \citep{de-Pontieu:2007hb}, followed by a rapid fading in 
chromospheric lines without any observed downfall. 
Spicules of type~{\sc ii} seen in the Ca~{\sc ii} band of Hinode show 
fading on time-scales of the order of a few tens 
of seconds \citep{de-Pontieu:2007kl}. The disk counterparts 
to this class of spicules are denominated ``Rapid Blue-shifted Events'' 
(RBEs) \citep{Langangen:2008fj,Rouppe-van-der-Voort:2009ul}. 
These show strong Doppler blue shifts in lines formed in the middle to 
upper chromosphere. \citet{De-Pontieu:2009yf} linked the RBEs with asymmetries 
in transition region and coronal spectral line profiles.
The increase in line broadening during the lifetime of RBE suggest that 
they are heated to at least transition region temperatures 
\citep{de-Pontieu:2007kl,Rouppe-van-der-Voort:2009ul}. 
Another intriguing observation is that type~{\sc ii} spicules often also
appear in regions that seem unipolar  \citep{McIntosh:2009yf}.

The latter observation complicates explanations of the type~{\sc ii}
phenomena by magnetic reconnection, which seems a likely candidate for
other types of coronal jets. For example, the pioneering 
2D simulations done by \citet{Yokoyama:1995uq,Yokoyama:1996kx}, or more 
recent simulations done by \citet{Nishizuka:2008zl} shows that 
emerging magnetic flux reconnects with an open ambient magnetic
field forming a large angle to the emerging field. Such reconnection
produces strong velocities and a jet, with a large energy release. 
3D experiments of jet formation in a horizontally 
magnetized atmosphere as consequence of a reconnection trigger, 
where the flux emergence is a prime candidate have been done by 
\citet{Archontis:2005rx,Galsgaard:2007mz}. 
\citet{Moreno-Insertis:2008ms} expanded the \citet{Yokoyama:1995uq} 
model to 3D, where the ambient field is open and connects into the corona. 

In this paper discussion is confined to type~{\sc ii} spicules, since 3D models of
the first type have been described previously by \citet{Martinez-Sykora:2009kl}.
We analyze a jet-like feature that we find in a realistic 3D-MHD simulation 
that shares many of the characteristics of the observed type~{\sc ii} 
spicules. The simulation includes non-grey and non-LTE radiative transfer 
with scattering, and thermal conduction along the magnetic field lines. 
The model spans the upper layers of the convection zone to the lower corona. 
In \S~\ref{sec:equations} the numerical methods are briefly described,
while the initial model and boundary conditions employed are described
in \S~\ref{sec:condition}. In \S~\ref{sec:st}, we describe the
structure of the spicule-like jet which is 
found in the model. We succeed in isolating the mechanism far enough to 
find the initial configuration in the photosphere that starts
the chain of events leading to the jet
(see \S~\ref{sec:birth}). The physics, dynamics, and the detailed evolution 
of the formation mechanism are described in \S~\ref{sec:source}. The fading 
of the jet from the chromosphere is discussed in \S~\ref{sec:fading}.
In \S~\ref{sec:consec} the mass and enthalpy flux associated 
with the jet is analyzed. Conclusions are presented in \S~\ref{sec:conclusions}.

\section{Equations and Numerical Method}
\label{sec:equations}

The MHD equations are solved in our physical domain using the 
{\it Oslo Stagger Code} (OSC). 
The numerical methods have been described in detail in 
 \citet{Dorch:1998db,Mackay+Galsgaard2001,paper1,Martinez-Sykora:2009rw}.
In short, the code functions as follows: A sixth order accurate 
method is used for determining the partial spatial derivatives. 
In instances where variables are needed at positions other than their 
defined grid positions a fifth order interpolation scheme is used. 
The equations are stepped forward in time using the explicit third order 
predictor-corrector procedure described by \citet{Hyman1979}, 
modified for variable length time steps. In order to suppress numerical 
noise, high-order artificial diffusion is added both in the forms of 
a viscosity and in the form of a magnetic diffusivity.

The radiative flux divergence from the photosphere and lower chromosphere 
is obtained by angle and wavelength integration of the transport equation 
assuming isotropic opacities and emissivities. The transport equation 
assumes that opacities are in LTE using four group mean opacities to cover
the entire spectrum \citep{Nordlund1982}. This is done by formulating the 
transfer equation for each of the four bins, calculating a 
mean source function in each bin. These source functions contain an 
approximate coherent scattering term and an exact contribution from thermal
emissivity. The resulting 3D scattering problems are solved by iteration, 
based on one-ray approximation in the angle integral for the mean intensity,
a method developed by \citet{Skartlien2000}.

In the mid and upper chromosphere the OSC code includes non-LTE radiative 
losses from tabulated hydrogen continua, hydrogen lines, and lines from singly 
ionized calcium as functions of temperature and column mass. These radiative 
losses depend of the computed non-LTE escape probability as a function 
of column mass and are based on a 1D dynamical chromospheric model in which 
the radiative losses are computed in detail 
(Carlsson \& Stein 1992, 1994, 1997, 2002).
\nocite{Carlsson:1992kl}
\nocite{Carlsson+Stein1994}
\nocite{Carlsson:1997tg}
\nocite{Carlsson:2002wl}

For the upper chromosphere and corona we assume optically thin radiative losses.
The optically thin radiative loss function based on the coronal
approximation and atomic data collected in the HAO spectral diagnostics package 
\citep{HAO_Diaper1994} is based on the elements hydrogen, helium, carbon, 
oxygen, neon and iron.

The model includes thermal conduction along the magnetic field. The conductive
term is split from the energy equation and is discretized using the 
Crank-Nicholson method. The resulting operator is implicit and is 
solved using a multi-grid solver \citep[see][]{paper1}.

The energy dissipated by Joule heating is given by $Q_{J}={\bf E \cdot J}$  
where the electric field ${\bf E}$ is calculated from the current ${\bf J}$,
taking into account high-order artificial resistivity. The resistivity
is computed using a hyper-diffusion operator 
\citep[see for details][]{Martinez-Sykora:2009rw}. This 
entails that the Joule heating is set proportional to the current squared 
times a factor that becomes large (of order 10) when magnetic field 
gradients are large, and is unity otherwise. 

The hyper-diffusion scheme used in the code makes it possible to run the code 
with a global diffusivity that is at least a factor 10 smaller than without. 
Unfortunately, the use of hyper-diffusivity also makes it impossible to give a 
single value of the Reynolds and magnetic Reynolds numbers for a given simulation. 
Even the effective magnetic Reynolds number is neither uniform nor easy to estimate 
as we do not have any clear ``magnetic boundary layer'' in the chromosphere, which is 
useful in estimate the effective magnetic Reynolds number \citep[see][]{emonet1998}. 
However, we stress that even if the Reynolds number is small, the
simulation does show velocities close to $100$~km~s$^{-1}$. Another method to
estimate the effective Reynolds number is to use our knowledge of the 
Joule heating ($\eta J^2$) from the simulation; we calculate $J$
from the curl of the magnetic field; and we 
can assume that the typical length is the length of the jet (say 4~Mm) and note that the 
Alfv{\'e}n speed in the chromosphere is of order 500~km~s$^{-1}$. Using these numbers give an 
effective magnetic Reynolds number of approximately 25. Note that the diffusivity used 
in this estimate is calculated in the vicinity of the largest Joule heating. In regions where 
the gradients and therefore dissipation is smaller, the diffusivity is also smaller, roughly 
a factor 8, and the Reynolds number is of order 200. We also note that the code described 
here can reproduce events with fast reconnection in 2D at a similar
spatial resolution as in this paper, an example 
of such processes are described in detail by \citet{Heggland:2009lr}
where chromospheric reconnection and the resulting jets are modeled. 

\section{Initial and boundary conditions}
\label{sec:condition}

\begin{figure}
  \includegraphics[width= 0.98\textwidth]{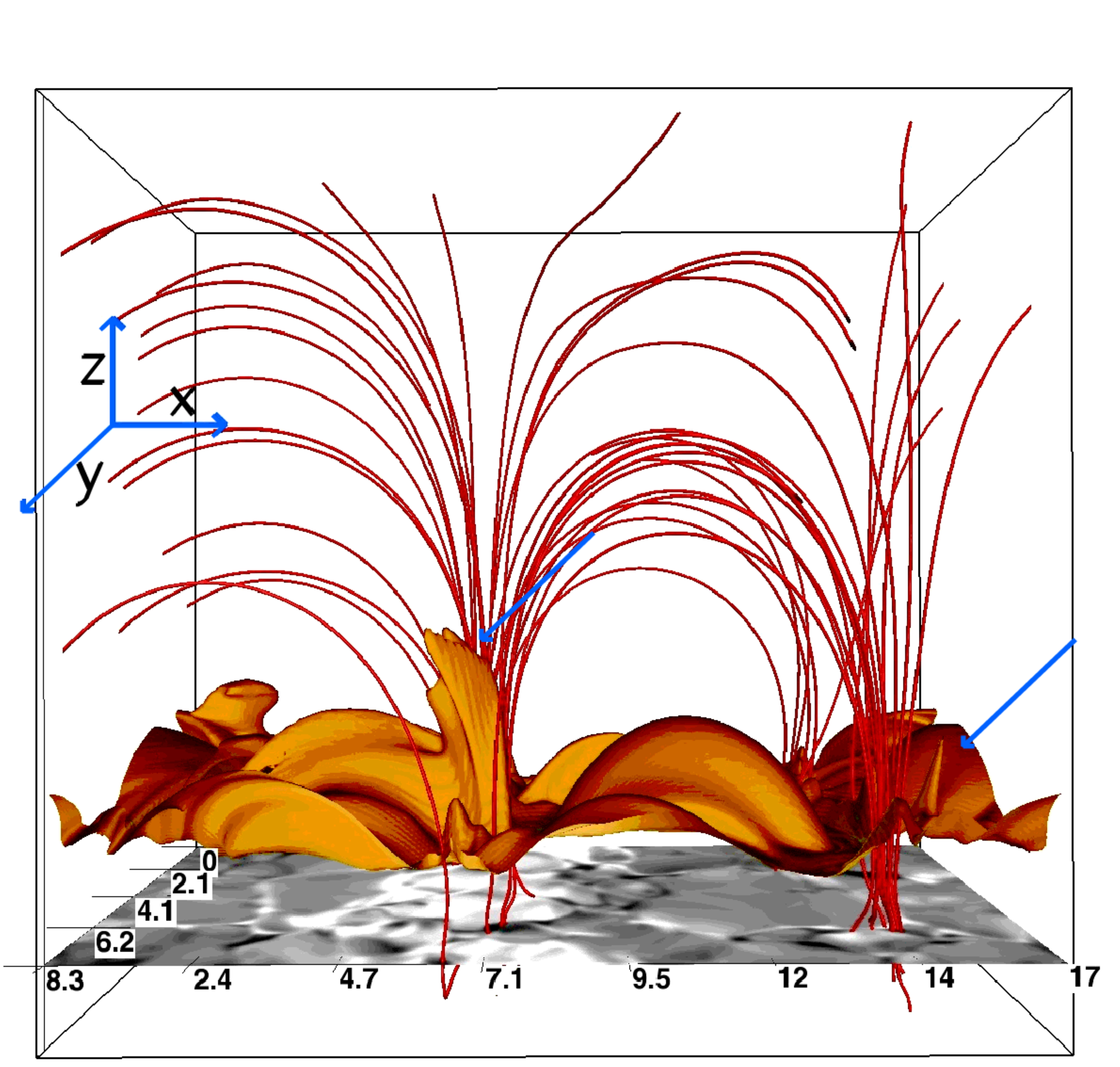}
  \caption{\label{fig:init} 
Magnetic field geometry and hydrodynamic configuration for the B1 simulation at 
$t=1850$~s: Field lines for the ambient pre-existing magnetic 
field (red), transition region surface at temperature $10^5$~K (orange isosurface), $B_y$  
magnetic field component in the photosphere (grey scale with white $210$~G and black $-210$~G). 
Note the location of the simulated spicule of type~{\sc ii} in the transition region 
isosurface (blue arrow at $x\approx 7$~Mm) and the region of flux emergence in the photosphere 
(where $B_y$ is white). An example of a simulated type~{\sc i} is visible to the right 
(blue arrow at $x \approx 14$~Mm).}
\end{figure}

The model discussed in this paper is described in detail in 
\citet{Martinez-Sykora:2009kl,Martinez-Sykora:2009rw}, where it is labeled 
as model ``B1''. 
The computational domain for this model stretches from the upper convection 
zone to the lower corona and is evaluated on a non-uniform grid of 
$256\times 128\times 160$ points spanning  $16\times 8\times 16$~Mm$^3$. 
The frame of reference for the model is chosen such that $x$ and $y$ are the 
horizontal directions, as shown in Fig.~\ref{fig:init}, which shows a 
selection of field lines, the $B_y$ component of the photospheric magnetic 
field, and the location of the transition region at time $t=1850$~s.
The grid is non-uniform in the vertical direction ($z$-axis) to ensure that the 
vertical resolution is good enough to resolve the photosphere and the 
transition region with a grid spacing of 32.5~km, while becoming larger at 
coronal heights where gradients are smaller and scale heights larger. At 
this resolution the model has been run for roughly 1 hour solar time. 

The initial model is seeded with magnetic field, which rapidly receives 
sufficient stress from photospheric motions to maintain coronal temperatures 
($T>500\,000$~K) in the upper part of the computational domain, in the 
same manner as first accomplished by \citet{Gudiksen+Nordlund2004}. 
The initial field was obtained by semi-randomly spreading $20-30$ 
positive and negative patches of vertical field at the bottom boundary some $1.5$~Mm below
the photosphere, 
then calculating and inserting the potential field arising from this 
distribution in the remainder of the domain. Stresses sufficient 
to maintain a minimal corona are built up by photospheric motions after 
roughly $20$~minutes solar time.  The model has an average unsigned field 
in the photosphere of $160$~G and is distributed in the photosphere 
in two bands of vertical field centered around roughly $x=7$~Mm and
$x=13$~Mm. In the corona this results in loop-shaped structures that 
stretch between these bands and are mainly oriented in the 
$x$-direction, as can be seen by following the red field lines shown 
in Fig.~\ref{fig:init}. Note that the time ($t=1850$~s) shown in the figure was chosen 
to coincide with the ejection of the spicule 
of type~{\sc ii} that is the subject of this paper. The jet is 
clearly visible in the isosurface that outlines the position of 
the transition region near $x=7$~Mm.

To start the events leading to the jets, we introduce a non-twisted
magnetic flux tube into the computational domain through 
its lower boundary, that lies some $1.5$~Mm below the 
photosphere, as described in detail by \citet{paper1}, section~3.2. 

Horizontal magnetic flux of strength $10^3$ G is injected in a band of 
$1.5$~Mm of diameter parallel to the $y$-axis and centered at $x=8$~Mm at the bottom boundary. Hence, the injected magnetic field is nearly perpendicular to the orientation of
the pre-existing ambient magnetic field outlined by the coronal loops that are seen
in Fig.~\ref{fig:init}. The details of the input parameters of the simulation 
can be found in \citet{Martinez-Sykora:2009rw}. 
The simulation has, therefore, on the one hand, a pre-existing field which interacts with granular convection 
similar to \citet{abbett2007,Isobe:2008lr}, and on the other hand, it has new emerging magnetic flux 
which is injected through the bottom boundary \citep{archontis2004,cheung2007,paper1} 
and interacts with the granular convection and the pre-existing magnetic field.  

\section{Results}
\label{sec:results}

In the various simulations we have run, and in particular in the simulation 
discussed here, a large number of spicule-like structures are found. 
Most of these resemble spicules of type~{\sc i}, as described by
\citet{Martinez-Sykora:2009kl}. However, in addition, we also find jet-like 
structures that resemble spicules of type~{\sc ii} \citep{Martinez-Sykora:2010vl}. 
These differ markedly in their physical characteristics  from the
pervasive type~{\sc i}.

Simulation B1 produced on the order of 100 spicules of type~{\sc i}. 
These spicules were found to result from upwardly propagating shocks
passing through the upper chromosphere and into the transition region and
corona. The shock-driven jets show up as narrow linear structures reaching
some few Mm above the ambient chromosphere, rising and falling with
velocities of order 30~km~s$^{-1}$, and having lifetimes of order
3-5~minutes. An example of this type of jet is seen on the right-hand side of
Fig.~\ref{fig:init}, (blue arrow near $x=14$~Mm) which appears like a narrow
spike in the transition region.  On the other hand, the simulation produced
two jets that resemble spicules of type~{\sc ii}, but only two of them; one of these is
shown in Fig.~\ref{fig:init} as a prominent feature in the transition
region (blue arrow near $x=7$~Mm).  In this latter type of jet,
chromospheric material is ejected far into the corona at very high
velocity and heated, reaching velocities up to 95~km~s$^{-1}$; 
the jet appears to fade without falling back towards the solar surface.

\subsection{Structure}\label{sec:st}

Both of the candidate spicules of type~{\sc ii} are located near the 
footpoints of coronal loops. Therefore, the field lines along which the 
jet is ejected are either open or, at least, penetrate far into the 
corona. Moreover, both jets appear near $x=7$~Mm (see blue arrow in Fig~\ref{fig:init}), 
close to locations where flux emergence is vigorous. 
In fact, we note that both jets first appear after the rising flux 
tube injected at the bottom boundary has emerged to and, in part, has crossed 
the photosphere. 

\begin{figure}
  \includegraphics[width=0.48\textwidth]{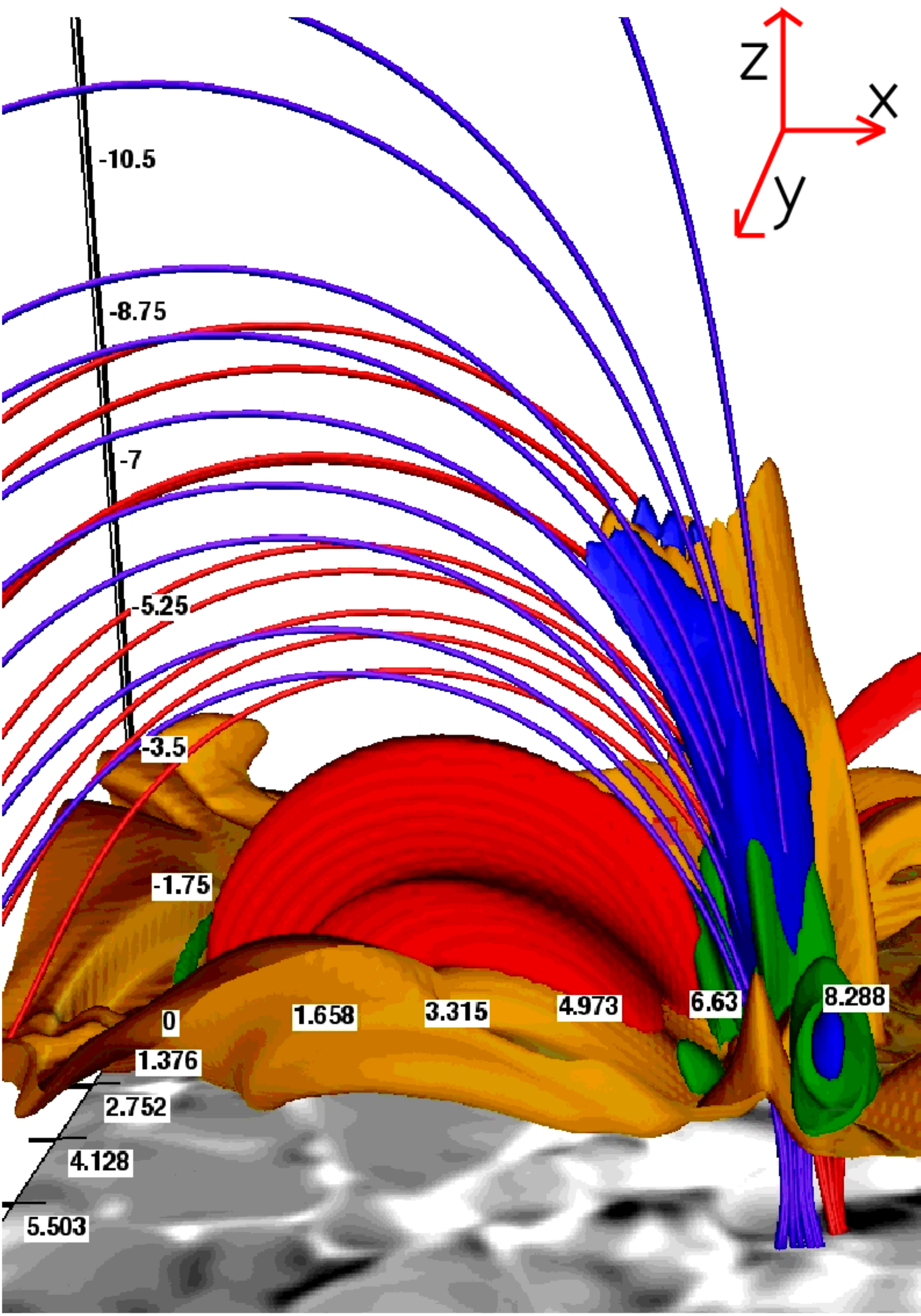}
  \caption{\label{fig:field} 
The structure of the type~{\sc ii} spicule and the related hot loop at $t=1850$~s, in terms of the  
temperature at $1.3\,10^6$~K (red isosurface), the transition region at $10^5$~K 
(orange isosurface), upflowing velocity along
the field lines at $45$~km~s$^{-1}$ (blue isosurface),  and Joule 
heating at $1$~W/m$^3$ (green isosurface).
The field lines are drawn on opposite sides of the
magnetic discontinuity in red and purple. 
$B_y$ at the photosphere is shown with a grey-scale map with a 
range of $[-210,210]$~G. Except in the spicule itself the transition
region is on average some 2~Mm above the photosphere.} 
\end{figure}

\begin{figure}
  \includegraphics[width=0.98\textwidth]{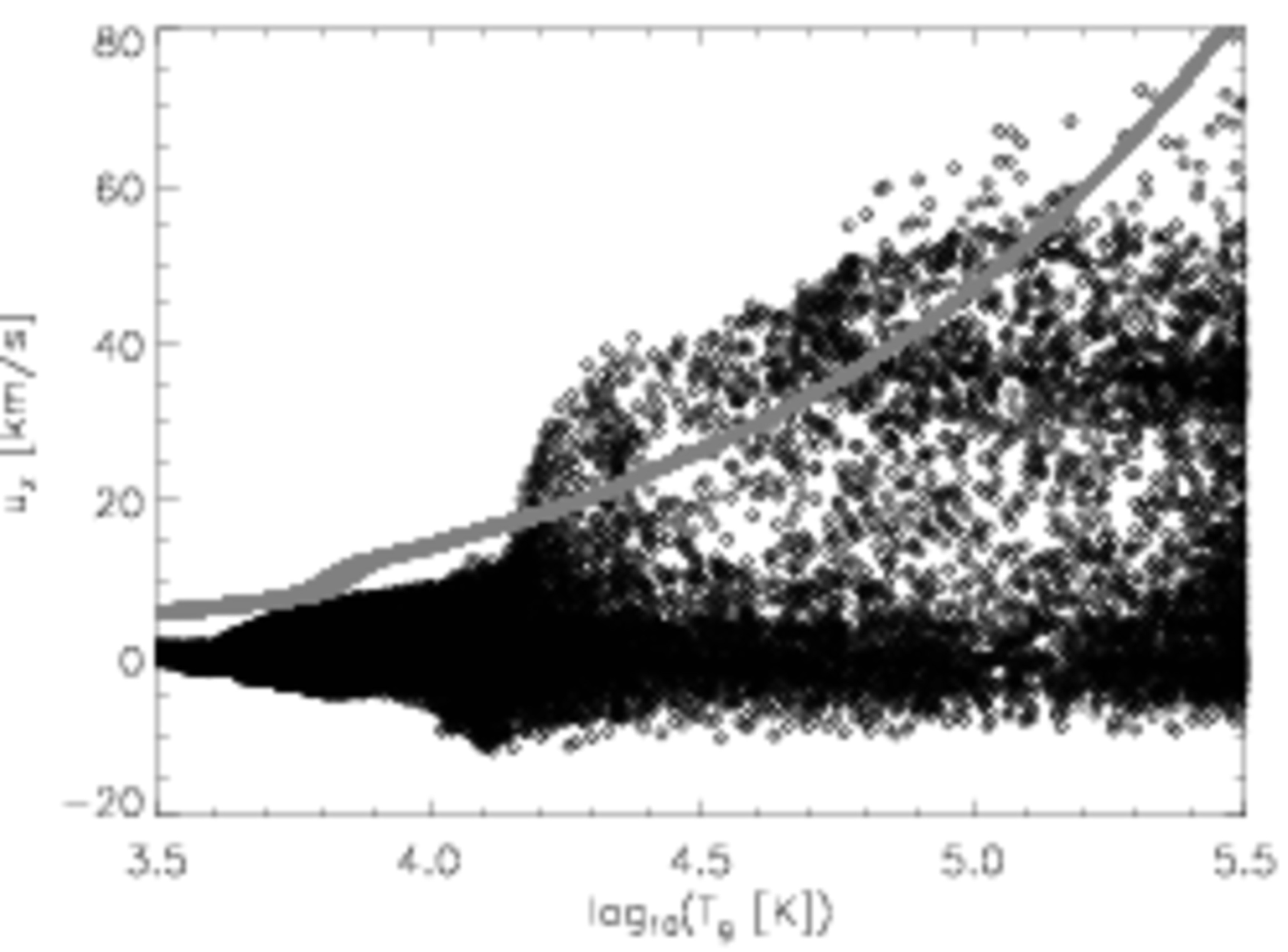}
  \caption{\label{fig:uztg_spic_015} 
  Distribution of vertical velocity $u_z$ as function of temperature
  $\log(T)$ in the vicinity of the spicule at time $t=1930$~s. Also shown with grey
  markers is the  distribution of the speed of sound
  $c_s=\sqrt{{\gamma P/\rho}}$ at the same time. The spicule plasma
  is supersonic and has temperatures in the range from $15\,000$~K
  to some few $10^5$~K at this time in the evolution of the spicule.}
\end{figure}

In Fig.~\ref{fig:field}, details of the type~{\sc ii} spicule structure are shown for 
the jet that is active at time $t=1850$~s in the B1 
simulation: in the figure, isosurfaces for the upflow velocity (blue), 
temperature (orange: transition region; red: hot corona), and  Joule heating 
per particle (green) are shown. The greyscale map towards the bottom of the
figure corresponds 
to the horizontal magnetic field strength in the photosphere which
clearly shows where flux emergence is vigorous.
In addition, the figure shows magnetic field lines close to the jet
structure. The jet is formed as  chromospheric material is ejected into 
the corona and heated, with velocities reaching up to $95$~km~s$^{-1}$. It is composed of 
many elements that together comprise a complicated phenomenon. In this paper, 
the spicule is identified with ejected material of chromospheric 
density and temperature, {\it i.e.} the jet-like structure seen in the 
orange temperature isosurface that appears next to the prominent spike seen 
in the blue velocity isosurface. The width of the jet
is roughly $500$~km and its maximal length is around $4$~Mm, measured from 
the initial position of the transition region. 
The velocity of the ejected material in the jet is aligned (or
very close to it) with the magnetic field and hence nearly vertical.
The vertical component of the velocity along a vertical axis passing 
through the jet shows a vertically bipolar structure with center in the 
middle-chromosphere; {\it i.e.} above $z=1.5$~Mm upflow velocities are 
found while below this height material is downflowing 
(for details see section~\ref{sec:source} and Fig.~\ref{fig:forc}). 
The maximum downflow velocity is of order $10$~km~s$^{-1}$. The 
upflowing ejected material has a distribution of velocities and 
temperatures: at 
its core it is cool, $15\,000$~K, and reaches velocities of 
$40$~km~s$^{-1}$, but the blue velocity isosurface of 
Fig.~\ref{fig:field} shows that there is warmer, faster upflow 
surrounding parts of 
the cool core with temperatures of a few $100\,000$~K, where the upflow 
velocity is higher than in the core, up to $95$~km~s$^{-1}$. 
These numbers fits fairly well
with the statistics of the RBE's done by \citep{Rouppe-van-der-Voort:2009ul}.
Further insight into the velocity structure can be found in 
Fig.~\ref{fig:uztg_spic_015} which shows the range of vertical velocities in
the volume surrounding and including the jet as a function of 
temperature, along with the speed of sound ($c_s=\sqrt{{\gamma P/\rho}}$). The 
velocities comprising the jet are largely supersonic, with Mach 
numbers of order two for the lowest temperature gas. 

Close to the base of the jet we find a region where the Joule heating 
per particle is large, as shown with the green isosurface in 
Fig.~\ref{fig:field}. Joule heating occurs along the entire length of
the jet, but is strongest in the first few megameters near the
base. As a result of Joule heating, the nearby corona is heated, and hot 
loops with temperatures up to 1.6~MK are produced with temperature much greater 
than that found in the ambient corona.
These hot loops are apparent through the red temperature isosurface. 
Note that the hot loops have one footpoint near the base of the jet. 
The hot loops consist of low-density material, and are associated with, but separate
from, the jet proper. 

It is interesting to note that the magnetic field is unipolar in the region 
around the modeled jet; this is also the case for many observed spicules 
of type~{\sc ii} \citep{McIntosh:2009yf}. Even so,
we find several regions with large gradients in the magnetic field 
in the chromosphere near the jet. 
More specifically, these discontinuities are found to be located inside a 
volume $2\times 2\times 3$~Mm$^3$ at $[x,y,z]=[3:7,5:7,1.5:4.5]$~Mm.  
The region of  magnetic discontinuities ({\it i.e} large magnetic
field gradients or current sheets) is
aligned  with the $x$-axis. These gradients show a magnetic field topology that is
similar to rotational discontinuities, with small inclinations between
the field lines on either side of the gradient. 
These gradients produce the strong Joule heating at the footpoints shown with 
the green isosurface in Fig~\ref{fig:field}.
The magnetic field configuration close to the jet is complex,
and contains many locations with strong magnetic gradients. 
The jet is therefore not solely associated with a simple discontinuity, but rather 
with several locations of large electric currents. 

It is important to note that the velocities in the upper part of the 
jet ($z > 1.5$~Mm) are directed upward during the entire 
evolution. No downflows are found in the jet at any 
point in time, in contrast to the simulated spicules of type~{\sc i}, 
in which both up and downflows are found. 
The rapid fading of the jet at the end of its evolution is described 
in detail in section~\ref{sec:fading}. A movie of the jet evolution is
provided as on-line material accompanying this paper. %% The online material!!

\subsection{The photospheric precursor of the jet}\label{sec:birth}

A detailed description of the setup and evolution of the flux emergence 
in this simulation have been given in detail by \citet{paper1,Martinez-Sykora:2009rw}. 
\citet{Martinez-Sykora:2009rw} demonstrated 
that flux emergence causes the magnetic field to expand into the corona, 
and that this expansion produces discontinuities in the coronal field. 
We believe that this also applies to the regions of large magnetic field gradient 
obtained in the present paper. 
These discontinuities are the source of the Joule heating which produces 
the hot loops, and have the indirect consequence of producing the jet
upflow, as described below.

\begin{figure}
\includegraphics[width=0.98\textwidth]{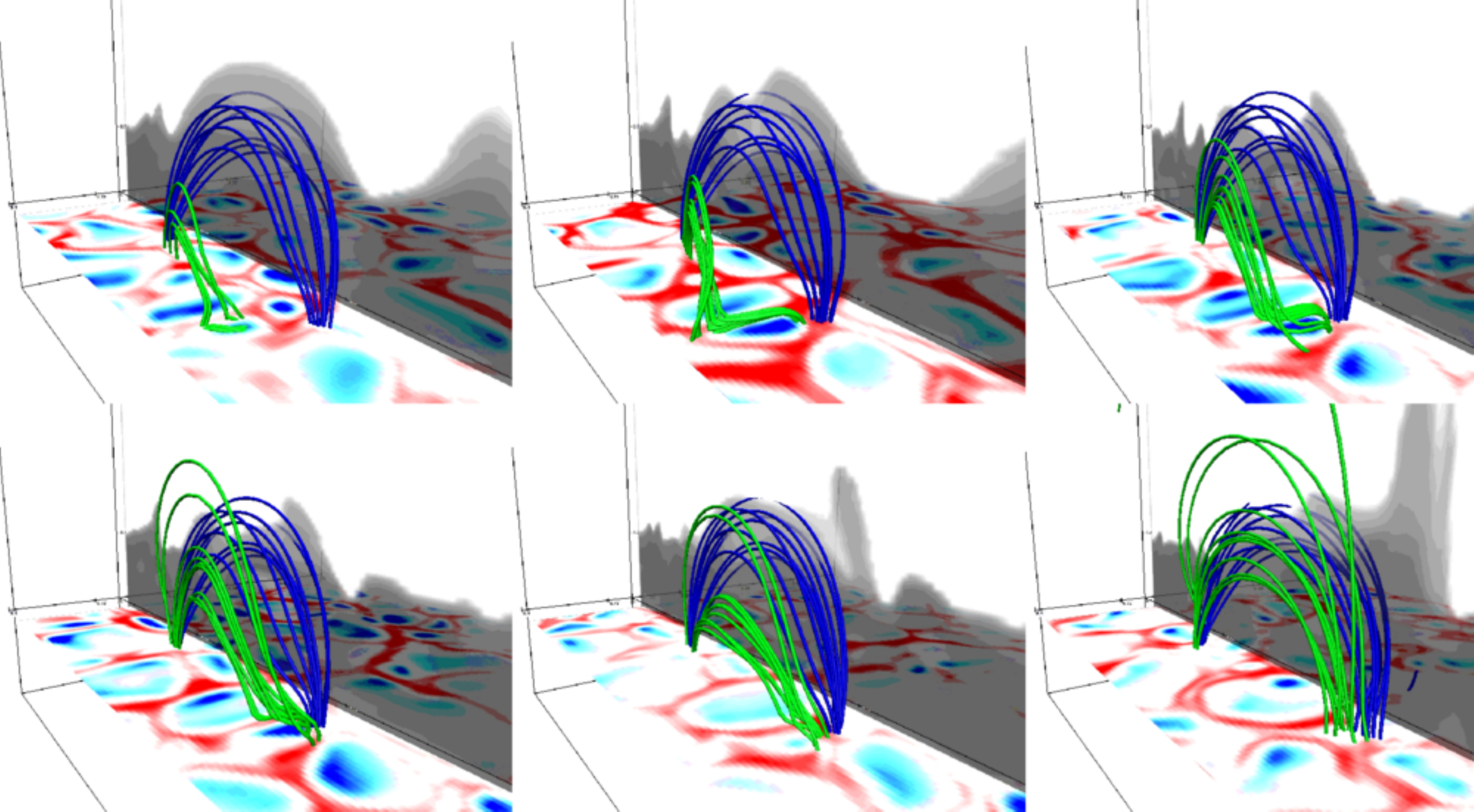}
  \caption{\label{fig:emerg3d} Horizontal magnetic field emerges through  
  the photosphere in granules and accumulates in the intergranular lanes. 
  The red-blue surface at 
  the photosphere shows the vertical velocity (red is downflow and 
  blue is upflow with range [-2,2]~km~s$^{-1}$). 
  The grey-scale surface shows the temperature in logarithmic scale in the 
  $xz$-plane at $y=6$~Mm. The 
  green magnetic field lines follows the bulk of plasma that emerges from the
  photosphere and expands into the corona. The blue field lines are taken at the same
  region in all timesteps near the upper chromosphere.  The time series is 
  t=[850,1050,1200,1300,1550,1900]~s from left to right and top to bottom.}
   \end{figure}

We have located emergence events in individual granules at the photosphere
that are likely precursors of the jet described in the other
sections. The sequence of images shown
in Fig.~\ref{fig:emerg3d} shows one such event. In the six panels comprising the figure
we illustrate the emergence and gradual blending of a set of field lines (green
lines) with a set of field lines representing the pre-existing field in the vicinity of
the jet (blue lines). Also shown is the velocity field in the
photosphere, i.e. at height $z=0$~Mm, and the logarithm of the
temperature in an $x-z$ plane placed at $y=6$~Mm (this is the same
plane used to describe the jet in Fig.~\ref{fig:forc}). In the upper
left panel, at $t=850$~s, the emerging field within a granule (see the horizontal 
component in the photosphere of the green lines) shows one footpoint 
connected to the convection zone, while the other footpoint is linked somewhere
below the photosphere to field lines that expand into the 
chromosphere, forming loops that are oriented at an angle to the
pre-existing field. In the top center panel, at $t=1050$~s, these field lines
are seen to connect to newly emerged horizontal field oriented in the
$y$-direction with footpoints near the pre-existing field foot points.
As the emerging field continues to rise into the chromosphere, shown
in the top right and bottom left panels at $t=1200$~s and $t=1300$~s
respectively, the strong hairpin curve in the emerging field is seen
to become shallower as the field straightens and the field lines begin 
to blend with the pre-existing field lines. In the bottom
center panel, at $t=1550$~s, the hairpin has largely disappeared 
and at the same time we see the first evidence of the jet in the
temperature plot to the right and above the displayed field lines. 
The bottom right panel, at $t=1900$~s, shows the
emerging field lines expanding and interacting with the pre-existing
field and that cool gas has been ejected into the corona to the right
of the system of field lines shown.

The evolution of the field lines shows they presumably have suffered
reconnection before appearing at the surface: while in the granule interior,
emerging field lines seem to have reconnected with the pre-existing field
resulting in the hairpin configuration of the emerging field lines. 
 {\it Hybrid} field lines (drawn in green in Fig.~\ref{fig:emerg3d})
 result, linking the emerging system in the granule interior with the
 pre-existing ambient coronal field, and with the high-curvature
stretch near the location where the reconnection is likely to have
occurred.  The high-curvature portion of these
field lines has a sharp hairpin shape which is difficult to reconcile with
having resulted from processes other than reconnection, such as {\it e.g.},
the deformation caused by the granular flows.  This early reconnection may
take place at different heights in the granule interior but in any case in
high plasma-$\beta$ ($1$~Mm below the photosphere) and it therefore does
not have the spectacular 
consequences often associated with low plasma-$\beta$ reconnection;
{\it ie.} no jets or flows. This kind of
process might occur not only in the quiet sun, network or internetwork, 
but also in plage, where the field is mostly unipolar. Since our
primary interest is to understand the consequences of these severely deformed field
lines as they emerge into the chromosphere, we will not study the
sub-surface reconnection in detail, only mention that this (or similar
processes) can produce complex magnetic field configurations in the
high plasma-$\beta$ part of the atmosphere that can survive emergence into
the chromosphere. 

Photospheric footpoint motions and high curvature lead to large magnetic field gradients 
and eventually a reorientation of the field in the chromospheric and
coronal layers which is the ultimate source of the jet described in
detail in section~\ref{sec:source}. 
As mentioned above, the emerged field lines and the
pre-existing field form an angle to each other and a geometry similar
to a rotational discontinuity, but the change in the orientation is small. 
Though we see Joule heating, such a small inclination may not be large enough to produce
dynamically significant reconnection in the chromosphere.

In the simulation, only a few granules show field configurations of sufficient complexity to
lead to discontinuities in the overlying
chromosphere. Hence, only a few jets of the type discussed here are formed
and launched during the entire numerical experiment. 

\subsection{Source and connectivity}\label{sec:source}

\begin{figure}
  \includegraphics[width=0.98\textwidth]{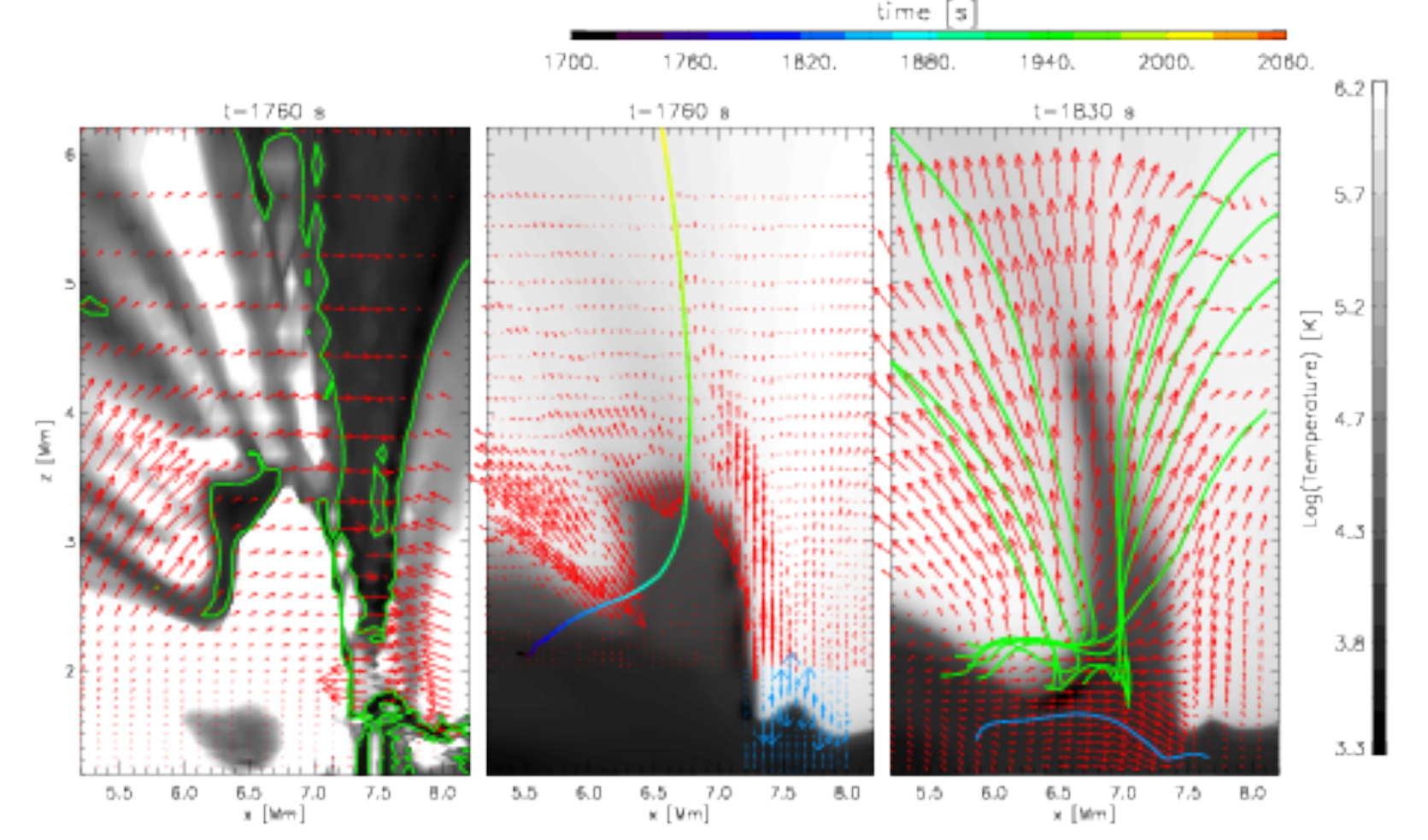} 
  \caption{\label{fig:forc} The spicule is formed as it is squeezed by the Lorentz force and 
  material is ejected by the resulting pressure gradient. The ratio of Lorentz force to pressure gradient
  $|{\bf j}\times {\bf B}|/|\nabla P|]$ (grey scale left panel, black
  where the gas pressure gradient dominates), the green line shows
  where the ratio is equal to $1$. The temperature on a logarithmic scale at time 
  $1760$~s, (grey scale middle and right panel) before the spicule ejection and at time 
  $1830$~s (right panel). 
 All panels are shown at the plane $y=6.02$~Mm. In the left panel the vectors show 
 $({\bf j}\times{\bf B}/\rho)$ (red),  in the middle panel $(\nabla
 P/\rho)_{||}$ (red and blue, the scale of the red vectors is half the
 value of the blue) and in the right panel the velocity field (red). 
Particle trajectories are shown in green (upflow) and blue (downflow) in the right panel. The 
particle trajectory shown as a function of time with the color-scheme in the middle panel is
analyzed in detail in Fig.~\ref{fig:urtrac}.}
\end{figure}

\begin{figure}
  \includegraphics[width=0.98\textwidth]{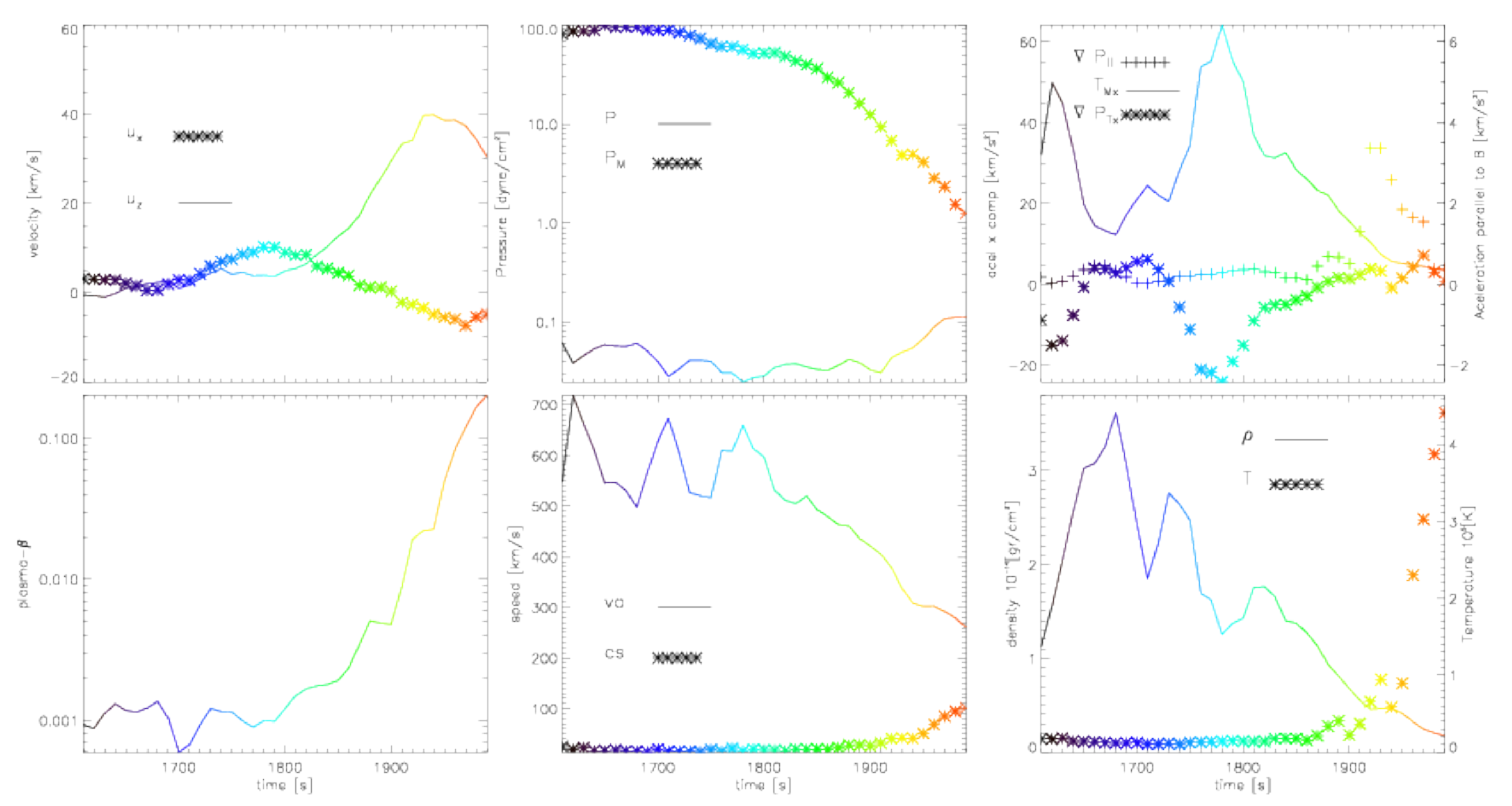} 
  \caption{\label{fig:urtrac} Different quantities evolution as a function of time of the particle which 
  path is shown in the middle panel of Fig.~\ref{fig:forc}. The vertical (solid line) and $x$-component 
  (asterisk symbols) of the velocity are shown in the top-left panel. Plasma-$\beta$ is shown in 
  the bottom-left panel. The gas pressure (solid line) and magnetic pressure (asterisk symbols) 
  are shown in the top-middle panel. Alfv\'en (solid line) and sound speed (asterisk symbols) are
  shown in the bottom panel. The acceleration in the 
  $x$-direction due to the magnetic tension ($(B\nabla) B/(2\rho \mu)$, solid line, left vertical axis), 
  due to the gradient of the magnetic pressure (asterisk symbols, left vertical axis), and due to the pressure 
  gradient along the field of lines (plus symbols, right vertical axis) are shown in the top-right panel. 
  Density (solid line) and temperature (asterisk symbols) are shown in the bottom-right panel.
  The color scheme corresponds to the time and is the same as in Fig.~\ref{fig:forc}.}
\end{figure}

The large magnetic field gradients in the chromosphere 
described in section~\ref{sec:st} do not 
produce an upflow jet, but rather horizontal flow, and the acceleration 
associated with the discontinuity is small. 
How then can jets that pull chromospheric material up to 
$7$~Mm above the photosphere at velocities $>60$~km~s$^{-1}$ be
produced? In Fig.~\ref{fig:forc} and \ref{fig:urtrac} we show various
aspects of the force balance in the vicinity of the jet that occurs
near $t=1850$~s. 
The arrows in Fig.~\ref{fig:forc} correspond to the Lorentz acceleration (left panel), the 
pressure gradient acceleration along the field lines before the jet 
is launched (middle panel) and the velocity field when the jet is launched (right panel). 
The grey scale map in the left panel compares the absolute value of
the Lorentz force and the pressure gradient, 
just before the jet is formed at $t=1760$~s (white means that the Lorentz force is
larger than the gradient of pressure). The temperature is shown with the grey-scale map 
in the middle panel at $t=1760$~s and in the right panel at $t=1830$~s, when the jet has 
evolved. These plots imply that the
chromospheric material actually is thrown into the corona by a 
gas pressure gradient, after being
initially accelerated horizontally by the Lorentz force.
 
That the cold material found at great heights is indeed the result of an ejection from 
the chromosphere can be seen by studying the particle trajectories of material 
composing the jet, shown as green, blue, and multicolored lines in the
middle and right panels of Fig.~\ref{fig:forc}. Particles clearly
start in the chromosphere before being ejected to coronal heights.
The velocity field after the jet is formed, at $t=1830$~s, is shown in the right 
panel of Fig.~\ref{fig:forc}. The velocity has a positive horizontal gradient in the 
chromosphere on the left side of the cold jet that 
has formed. The horizontal $x$-component of the velocity also increases with time at 
chromospheric heights in locations near the jet.
Figure~\ref{fig:urtrac} shows the vertical and $x$-component of the velocity for the trajectory 
shown in the middle panel of Fig.~\ref{fig:forc}. The particle
horizontal velocity increases with time, in the interval from 
1700 (purple) to 1800~s (blueish). Then, the particle is deflected and $u_z$ increases 
considerably, up to 40~km~s$^{-1}$. The particle evolves under
chromospheric conditions during this time material and the temperature
remains below $10^5$~K (see right-bottom panel) until $t=1930$~s. 
Note that Fig.~\ref{fig:uztg_spic_015} shows that there are many particles ejected with velocities
of the order of 40~km~s$^{-1}$ and temperature of the order $10^4$~K, as well as particles 
with 60~km~s$^{-1}$ and temperature of the order of $10^5$~K. 
Observe that plasma-$\beta$ (left-bottom panel) is below $10^{-2}$ when the particle is 
accelerated along the $x$-direction, after which 
plasma-$\beta$ increases to $0.1$. We can also see the large difference 
between the magnetic pressure and the gas pressure in the middle-top panel. 
Therefore, the Alfv\'en speed is of the order of 500~km~s$^{-1}$ 
and the sound speed around 20~km~s$^{-1}$ (see middle-bottom panel), 
{\it i.e.} the plasma is highly magnetized and the horizontal force
balance roughly perpendicular to the field is completely dominated by
the Lorentz force.

In order to find out what drives to the plasma to be squeezed in and
later ejected from the chromosphere, 
examine the left panel of Fig~\ref{fig:forc}: there are two regions where the  
Lorentz acceleration is larger than the pressure gradient, 
one inside the cold jet material and the other one $0.5$~Mm further to the right
($x>7.7$~Mm). The plasma is 
squeezed between these two regions. The Lorentz force located to the left side and inside 
the jet ($x<7.5$~Mm) is produced by
the strong magnetic field gradients, where the field lines show a
small jump in the magnetic field line orientation. This type of gradient  
produces a current that flows in the $y$-direction. The magnetic field
is mainly vertical, oriented along $z$, and 
thus a strong Lorentz force results that points horizontally in the $x$-direction as 
shown by the red vectors inside the white area delineating the jet in the left panel 
of Fig.~\ref{fig:forc}. Making a decomposition between the magnetic tension and the gradient 
of magnetic pressure along the $x$-direction, we can see in the top-right panel of Fig.~\ref{fig:urtrac} 
that the particle suffers a strong magnetic tension in the $x$-direction. This panel shows the 
time evolution the magnetic tension and the total (gas and magnetic) pressure gradient 
in the horizontal direction for a specific particle (see the trajectory in the middle 
panel of Fig.~\ref{fig:forc}), This magnetic tension, 
which comes from the expansion of the emerging field lines, is the force which 
pulls the plasma to the right and squeezes it. 

Thus, the magnetic tension component of the Lorentz force is the active agent that 
moves the plasma from left to right in the $x$-direction. The Lorentz acceleration 
is by far the most important component in the region where the particles move horizontally, 
to the left of, and inside the jet. Clearly, the magnetic tension is the force that squeezes the 
plasma and increases the pressure. We want to clarify that the 
agent of the magnetic tension in the chromosphere is due to the 
expansion of the emerging field lines. We could not find any clear
reconnection, nor tangential discontinuities, i.e. the largest
inclination between the emergent magnetic field and the ambient 
magnetic field lines are smaller than 10 degrees. Therefore, any 
subsequent reconnection with such variation in the angles between
the field lines can hardly be the source of the strong magnetic tension. 

Considering only the Lorentz acceleration, and integrating along the particle path
for a particle moving along its trajectory until it reaches the region where it is 
ejected upward gives a velocity that is supersonic. 
The resulting particle trajectory is horizontal and the particles moves a distance on the 
order of $2$~Mm, as shown with the green lines in the right panel of
Fig.~\ref{fig:forc}. In the actual modeled case we find that the cool jet plasma is moving
horizontally with nearly the speed of sound before it is deflected in
the vertical direction as shown in the upper left panel of Fig.~\ref{fig:urtrac}.

It is interesting to note that the Lorentz force squeezes the cold chromospheric 
plasma into a rather narrow structure. The flow above a specific vertical height is upward while 
the velocity is directed downward below this height (see right panel of Fig.~\ref{fig:forc}). 
However, what force deflects the flow in a vertical direction and pushes the plasma 
into the corona? The top right panel of Fig.~\ref{fig:urtrac} also
shows the gas pressure gradient parallel (almost vertical) to the magnetic
field, where there is no Lorentz force. Initially, the particle is accelerated 
horizontally by the Lorentz force and the horizontal velocity becomes nearly supersonic. 
However, when the particle gets closer to the region where the plasma is being squeezed,
both the gas pressure gradient and the magnetic pressure gradient act against this
flow. The gas pressure gradient becomes large
along the field lines, with no resistance to motion other than the gravity. In some 
places the pressure gradient acceleration reaches values nearly 10
times larger. For the particular particle track shown in
Fig~\ref{fig:urtrac} we find that the vertical velocity increases by
20~km~s$^{-1}$ between $t=1850$~s and $t=1910$~s, while in the same period
the temperature is remains below $30\,000$~K. Integrating ${(\nabla
P)_{\parallel}/\rho}-g$ over the same period gives 16~km~s$^{-1}$: 
most or all of the vertical velocity increase is due the
pressure gradient along the field and we conclude that
while the horizontal force balance is dominated by the
Lorentz force; the gas pressure gradient plays an important
role in turning the horizontal gas flow and especially in
accelerating the jet vertically. The jet has a complex 3D
structure, and the different forces, velocities, heating and
accelerations are not located in the same position, it is therefore
quite difficult to follow in only one particle all the processes
which happen in the spicules, but we note that already at time
1820 s the gradient of pressure is more than five times larger
than gravity.

On the right side of the jet, at the boundary between the cool chromospheric material of the jet 
and coronal temperatures, the horizontal flow reaches a ``wall'' where the pressure gradient 
takes over, and the material is accelerated vertically (see dark region in the right side of 
the jet in the left panel of Fig.~\ref{fig:forc}). This pressure gradient is large enough 
to push the flow up to heights 7~Mm above the photosphere. The wall on the right side
of the jet that forces an increase in the pressure is due to the Lorentz force 
(see vectors at the right side of the jet in the left panel of Fig.~\ref{fig:forc}). 
This wall is not of the same nature as the Lorentz force on the left and inner side of the 
jet, as there is no horizontal flow nor large magnetic field gradients associated with it, 
but the hot plasma there is magnetically dominated ($\beta <<1$) and the vertical magnetic 
field has a much larger energy density than that contained in the plasma that is colliding 
with the magnetic field forming the wall.  
The wall concentrates the plasma on the right side of the jet and thus increases 
the pressure, producing a deflection of the flow. The jet slowly moves to the 
right with time (see middle and right panel of the Fig.~\ref{fig:forc}), along with the wall itself
and the region where the plasma is squeezed and forced to move vertically. 

That the horizontal flow of the jet is deflected by the pressure gradient 
is also illustrated with the vectors drawn in the middle panel of Fig.~\ref{fig:forc}. These 
vectors are over-plotted on an image of the  temperature structure in the vicinity of the 
jet just before the ejection. 
The vectors show that there are mainly two regions where the flow 
is strongly deflected by the pressure gradient; centered near $[x,z]=[6.0,2.9]$~Mm 
and near $[x,z]=[7.3,1.5]$~Mm, both aligned with the magnetic field. The leftmost of 
these regions is located above the transition region where, later in time, a hot loop
will appear. The other region lies to the right, and it is from this latter region 
the cool jet rises. Observe that the region of large pressure gradient is not 
concentrated in one small spot but stretches a long distance along the jet
and the hot loop. The plasma flow in the earlier stages (left and middle panels of 
Fig.~\ref{fig:forc}) of the event is more concentrated towards the right-hand side 
of the boundary between the chromospheric and coronal material than it is in later 
stages (right panel) which shows strong upflows in regions closer to the left side of 
the jet.  It is important to understand that even though the plasma-$\beta$ is low, the 
pressure gradient is aligned with the magnetic field lines and thus
dominates the dynamic along the field. Therefore, as mention above, the pressure gradient is free to
accelerate the plasma into the corona. We find that the magnetic
tension squeezing of the plasma increases the pressure in some regions by as much a
factor 10, as seen in the middle upper panel
of Fig~\ref{fig:urtrac}. Subsequently, note that the cool jet plasma is far from following 
the elementary adiabatic law  $P \propto T^{\gamma/(\gamma-1)}$: helium is ionized 
at some $10\,000$~K, increasing the number of
particles, on the other hand the ionization process costs energy and
the temperature remains fairly constant until ionization is complete.
Only when the gas is fully ionized does the temperature increase as rapidly 
as predicted in the elementary adiabatic case. While the gas pressure 
increase is of the order of 10 times
in the chromosphere the temperature increase is no more than a factor 1.4 during the most
vigorous part of the compression. The combination of large horizontal
inflow velocities, ionization/recombination processes and magnetic squeezing
results in a greater vertical velocity than would otherwise be
predicted \citep{Shibata:1982qy,Shibata:1982fk}.

The tension force that drives the horizontal motions is a consequence 
of the magnetic field that penetrates into the chromosphere. This field is a result of 
the reconfiguration in the photosphere due to previous reconnection and emergence. As 
described earlier strongly bent field lines expand into the chromosphere and butt against
the previously existing field. Note that, in the chromosphere the difference in 
orientation between the emerging field and the pre-existing field is small 
(less than 30 degrees).  The emerging field lines are not in equilibrium and penetrate 
high into the upper chromosphere (see Fig.~\ref{fig:emerg3d}). The
magnetic tension that squeezes 
the plasma and drives the jet results from this expansion of field lines that are
not well aligned with the previously existing ambient field. The current sheets that 
form between these flux systems are also the sites of enhanced Joule heating and 
potentially of reconnection. However, we have not found any clear evidence of 
reconnection at the chromospheric level, 
and since the relative angle between the magnetic field systems is small, it seems 
unlikely that eventual hidden reconnection dominates the jet dynamics.

\begin{figure}
  \includegraphics[width=0.98\textwidth]{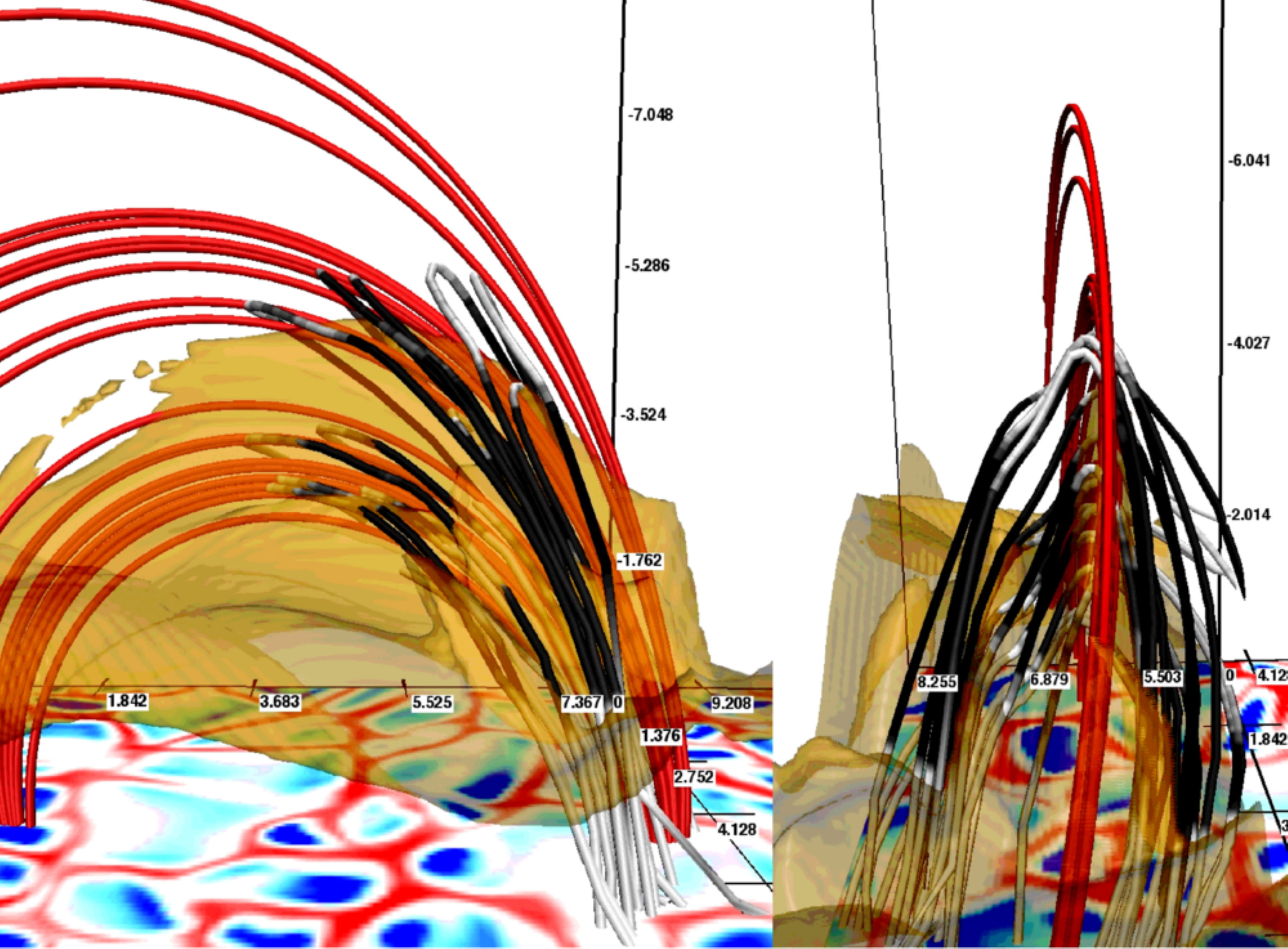} 
  \caption{\label{fig:conec} The structure and properties of the
    modeled spicule and its vicinity are connected with the
    configuration of the electric current and magnetic field. The
    field lines of the electric current are drawn with black-white colors, where 
  black and white indicate low and high Joule heating respectively. 
 The magnetic field are drawn with red lines. The transition region
 ($10\,^5$~K) is indicated using an orange semitransparent isosurface. The 
  photospheric vertical velocity is shown in a blue-red colored map with 
  range $[-2,2]$~km\,s$^{-1}$.}
\end{figure}

How do all these different structures; the upflows, the hot loop and the ejected 
chromospheric material, connect from the photosphere up to the 
corona? Is there any clear link between them? and why are they formed next 
to each other? 
The link between them lies with the orientation of the electric
current as compared to that of the magnetic field lines, as shown in 
Fig.~\ref{fig:conec}. The magnetic field configuration in combination with the current
field distributed spatially in the surroundings of the jet is
complex and difficult to visualize.  
In the vicinity of the footpoints of the ejected chromospheric material we find
a region of large Joule heating per particle (white 
colored region of the current lines near $(x,y,z)=(7,5:8,1.7)$~Mm). The current lines
in this region are mainly parallel to the magnetic field 
lines, this is where the gradient shows a topology of the field lines similar to a
rotational discontinuity with a small relative angle. Following the
current field lines higher up in the structure
the heating per particle is reduced (black region of the current lines, around $(x,y,z)=(6,5:8,2.3)$~Mm),
where the current lines are still parallel to the field lines.
Following the current field lines even further, the Lorentz force becomes 
important where ${\bf J}$ forms a large angle to ${\bf B}$, which is seen to
occur near the apex of the rendered current lines, where the current
changes direction and becomes perpendicular relative to the magnetic
field. Here the ${\bf J}$ is nearly perpendicular the current is large again (white small region 
at the top of the current field lines, around $(x,y,z)=(5,6.9,3.5)$~Mm).
Therefore, regions of large Lorentz force and regions of large Joule heating
per particle lie next to each other, but do not overlap.

\subsection{Fading away}\label{sec:fading}

Roughly one minute after the jet is ejected into the corona the cold material
contained in it starts to disappear. This material does not fall down again, instead it 
expands into the corona where it is heated to coronal temperatures. 
This process takes only a few tens of seconds. 

\begin{figure}
  \includegraphics[width=0.48\textwidth]{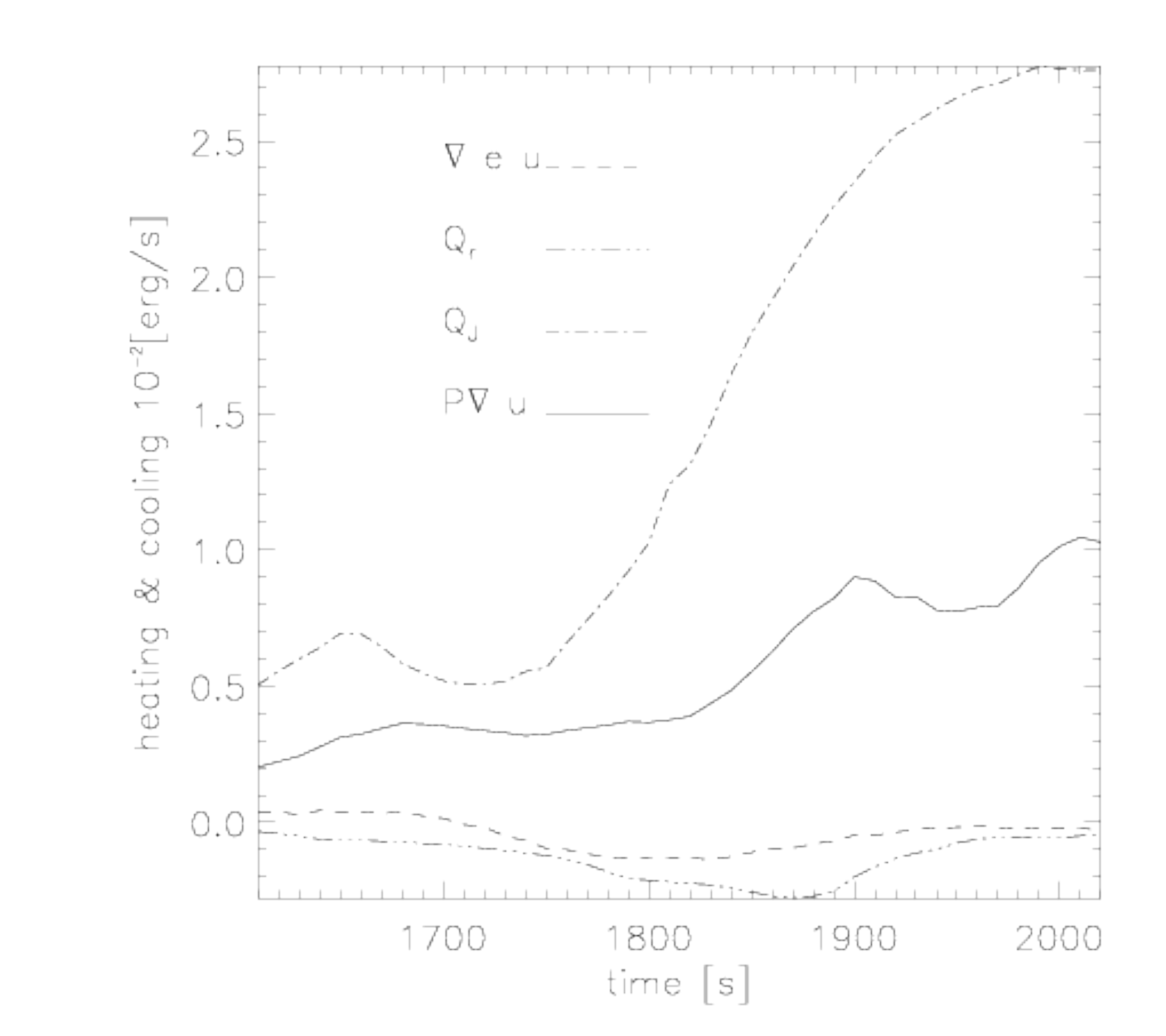} 
  \caption{\label{fig:energ} Energy balance is shown as a function of time at the footpoint of the 
  spicule. Mean value around the footpoint (centered at $(x,y,z)=(6.8,6.2,3)$~Mm and the mean
  is calculated within a volume of $1$~Mm$^3$) of the spicule of the advection of 
  energy ($\nabla e {\bf u}$, dashed line), radiative loss ($Q_r$, dot-dot-dot-dashed line), 
  Joule heating ($Q_J$, dot-dashed line) and compression heating ($P\nabla \bf{u}$, 
  solid line, note that this term is negative, energy lost) 
  are shown as a function of time. The plasma at the footpoint of the spicule 
  is heated by Joule heating.}
\end{figure}

The heating  associated with the jet is concentrated mostly 
in a region close to the footpoint of the cold jet. The different heating and cooling contributions  
are shown in Fig.~\ref{fig:energ} as a function of time. We calculated the mean value 
in a volume of roughly $1$~Mm$^3$ for each heating contribution. The volume 
is centered at the footpoint of the jet. The only component that 
contributes to the heating is the $Q_J$, since the compression heating
($P\nabla {\bf u}$) contributes mainly under expansion, {\it i.e.} it is cooling. 
Therefore, the important heating mechanism 
that heats the chromospheric material up to coronal temperature is the Joule heating. 
The Joule heating in this volume is large enough to heat all the mass that is injected into the 
corona ($4.9\,10^7$~kg, see section~\ref{sec:consec}) up to
temperatures of the order $10^{6}$~K. Assuming that we heat the gas in a 
cubic volume that spans $x=[6.3,7.2], y=[5.8,6.7]$ and $z=[3.54,2.54]$~Mm 
from chromospheric to coronal temperatures 
($T\approx10^6$~K and $\rho \approx 2\,10^{-14}$ gr~cm$^{-3}$) we estimate that this requires 
$(3/2)k_b(\rho/m) T \approx 2.5\,10^{24}$~ergs. Comparing this to an average Joule heating 
in the same volume measured to be $2.6\,10^{22}$ erg~s$^{-1}$, we find that Joule heating contributes 
$2.6\, 10^{24}$ ergs during the 100~s of the jet lifetime. 
The heating comes from the  
patches of Joule heating (see Fig.~\ref{fig:field}) that both produce the hot loop 
and heat the adjacent chromospheric material as it ascends. This injected energy also 
propagates along the field lines through conduction, once the plasma has reached 
sufficient temperature to make thermal conduction efficient, raising the temperature 
even in upflowing material far from the heating site. 

In the earlier stages of the ejection the density is high and the Joule 
heating per particle is  low. However, 
as the density of the cool chromospheric material decreases during expansion,
the heating rate per particle increases rapidly and the temperature of
the material rises. 
The volume where Joule heating is large grows with the jet and is present until the 
discontinuities in the magnetic field are relaxed and the current
becomes insignificant.

As mentioned previously the acceleration along the magnetic field lines is due to the gas pressure gradient. This increase in pressure 
is due to squeezing of the chromospheric plasma rather than by Joule heating. We have checked that the entropy increase in the plasma being accelerated is comparatively small: the largest entropy increase is located to the sides of the ejected chromospheric material. It is there, rather than in the jet itself, that the Joule heating per particle is largest. 

\subsection{Mass and enthalpy flux associated with the jet}\label{sec:consec}

\begin{figure}
  \includegraphics[width=0.48\textwidth]{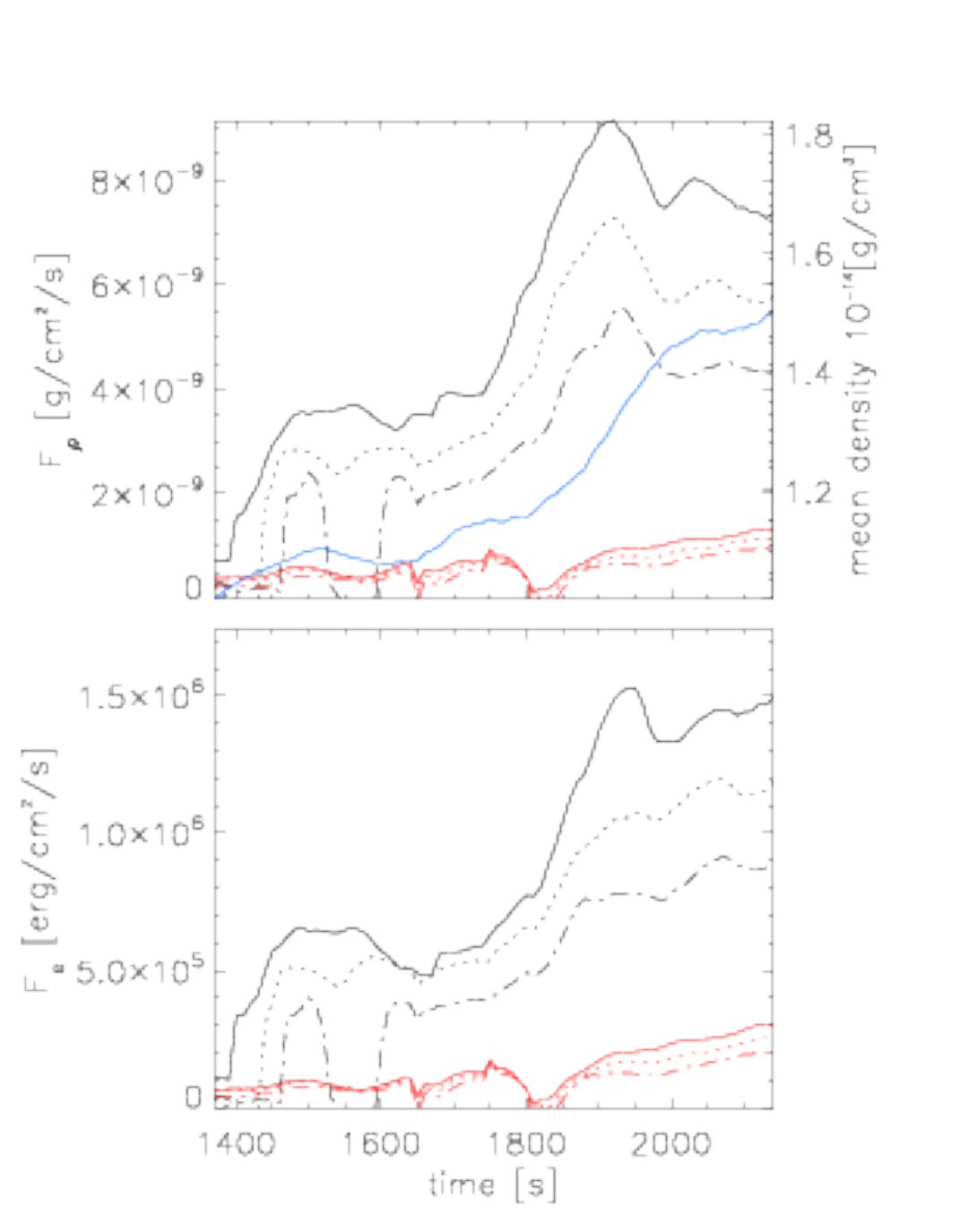}
  \caption{\label{fig:masslos} Chromospheric material is injected into the corona. 
  Mass flux (top panel) and energy flux (bottom panel) for the chromospheric 
  material of the jet structure (black lines) and in the 
    corona (red lines) as a function of time, shown at $z=[3.4,4.1,5]$~Mm 
  above the photosphere (solid, dotted, and dash-dotted lines respectively).
  The mean density in the corona ($T>10^5$~K) is shown in blue in the top panel.}
\end{figure}

Spicules inject mass, momentum, and energy  into the corona. The 
mass flux at fixed heights ($z=[3.4,4.1,5]$~Mm) is shown in the top panel of
Fig.~\ref{fig:masslos}. Red lines correspond to the total flux in regions with
coronal temperatures (more precisely: $T>10^5$~K) and black lines correspond to the chromospheric  
material of the jet (i.e., points with $T<10^5$~K and restricted to the
vicinity of the jet). 
The flux per unit of area is calculated as follows:

\begin{eqnarray}
F_{\rho}=\frac{\int_S(\rho \, u_z)ds}{\int_S ds} \;,\\
F_{e}=\frac{\int_S(u_z\, e)ds}{\int_S ds}\;,
\end{eqnarray}

\noindent where $\rho$ is the density, $u_z$ is the vertical velocity, and $e$ is the 
internal energy per unit volume. 
During the $700$~s displayed, the mass flux is always positive. Most of it is 
due to the injection and heating of chromospheric material from the
jet(s). We see two phases in the figure, coinciding with two jet
ejections: one starting at around $t=1400$~s, and the other, which has been the focus
of this paper, starting at time $1750$~s. 
The mass flux per unit area in the jet (black lines) reaches a value more than 4 times greater 
than the average inflow through high temperature points (red lines). The mass injected
into the corona by the jet along its $100$~s lifetime is on the order of $4.9\,10^7$~kg. 

In addition to mass, a large amount of thermal energy is also injected into the corona.
The enthalpy flux is shown in the bottom-panel of 
Fig.~\ref{fig:masslos} which follows the same color and line-type scheme as the top panel. 
The largest thermal energy inflow by the jet starts at time $1750$~s and 
is at least $3$ times greater than that 
entering the rest of the corona; the total thermal energy injection by 
the jet is about $8\,10^{24}$~erg during its lifetime \citep{De-Pontieu:2011lr}.

The mean density in the corona is shown in blue in the top-panel of Fig.~\ref{fig:masslos}. 
The mean density is calculated to be the total coronal mass divided by the coronal volume, 
defined as the volume in which the temperature is greater than $10^5$~K. 
The mean density in the corona increases considerably as the jet supplies mass into 
the high temperature regions of the upper atmosphere. 

\section{Discussion and conclusions}
\label{sec:conclusions}

We have studied the physics and dynamics of a plausible candidate for a type~{\sc ii} 
spicule model found in realistic 3D simulations. This candidate evolved
naturally as a consequence of the dynamics of the model. The mechanism is complex:
chromospheric material 
is ejected into the corona as plasma is pushed horizontally by a strong Lorentz force towards a
``wall'' of strong vertical magnetic field where plasma-$\beta < 1$. The wall causes a large 
increase in the pressure and thus deflects the plasma and forces it to
flow vertically along the magnetic field, most rapidly towards the low densities of the corona. 
The compression does not follow the elementary adiabatic process, because, mostly,
helium is ionized and the temperature remains fairly constant during the process.  
The Lorentz force which pushes the plasma results from large gradients 
in the field orientation that have accumulated in the upper chromosphere. These large gradients 
possess a current that is both horizontal and perpendicular to the magnetic field depending 
the spatial position and time. 

This mechanism cannot be identified with previous mechanisms described in the literature; 
it is different from the surges caused by propagating shocks \citep{Shibata:1982fk,Shibata:1982qy} 
or the spicules described by \citet{Hansteen+DePontieu2006,Heggland:2007jt,Martinez-Sykora:2009kl} among others. 
Such phenomena come from a shock which is formed in or crosses the chromosphere and 
pushes the transition region upwards. The mechanism discussed here is also different from 
the cases of collision of emerging plasma with a preexisting ambient magnetic field in the 
corona that lead to a large magnetic field discontinuity 
\citep[][among others]{Yokoyama:1996kx,Archontis:2005rx,Galsgaard:2007mz,Moreno-Insertis:2008ms,Nishizuka:2008zl}. 
Such simulations show that reconnection can produce a strong jet when the outflows from 
the reconnection site impinge upon the neighboring medium with subsequent squeezing 
of the plasma and acceleration along the field lines. At first sight the phenomena described 
in this paper could seem to have common features with that mechanism. However, in our 
simulation the reconnection, the collision and the plasma squeezing take place in quite 
different layers, from the convection zone to the chromosphere. The primary cause of the 
modeled type~{\sc ii} spicule in our model is not reconnection in the chromosphere, but 
rather the squeezing caused by the straightening and expansion of field lines rising from 
the photosphere. 

The ultimate source of the strong magnetic field gradients in the upper chromosphere
is the local emergence of magnetic flux that injects strongly stressed
magnetic fields into the photosphere just below the site of the ejected jet.
The main effect is the interaction between the flux emerging in the
granules and the intergranular ambient field:
the former suffer a reorientation while the lines 
connected to the intergranules do not. As they rise into the upper
layers of the atmosphere the former field lines show strong magnetic tension. 
These two system of field lines are the source of 
large gradients of the magnetic field orientation that arise higher in
the upper chromosphere with large current along the field of lines and 
Joule heating.

Joule heating near the footpoints of the ejected spicule initially produces  
a hot loop to the side of the ejection, where the heat is spread throughout the loop 
by thermal conduction. Later in time, this same heating also raises the
temperature of the cool spicule material to coronal values, in fact on a
fairly short time scale (on the order of a few tens of seconds) once the chromospheric
material has expanded into the corona.

Such processes have considerable effects on the modeled corona which potentially could be 
important. For instance, we observed a large injection of chromospheric 
material into the corona and an accompanying thermal energy flux sufficient to significantly 
contribute to heat the corona. In addition, large upflow velocities 
are introduced into the lower corona that carry a substantial kinetic energy flux. 

It is worth mentioning that the large magnetic field gradients needed to create
the conditions necessary to accelerate the spicules discussed in this paper 
could alternately be produced by other sources than flux emergence; for example
one could imagine that chromospheric dynamics in the high plasma-$\beta$ region of the 
atmosphere potentially could stress the magnetic field sufficiently. However, in the
model analyzed here we could only find candidate type~{\sc ii} spicules that were related with
flux emergence. This might be because the initial field configuration is severely 
simplified in the model compared to topologies found in the actual sun and/or
that the chromospheric model presented here could be improved, through
better spatial resolution or by adding other relevant physics beyond MHD. 
Thus, one important avenue of research is to analyze which magnetic field 
topologies are most likely to produce spicules of type~{\sc ii}: the quiet sun, plage,
or active regions. Alternatively, one could determine what other sources could lead to the development of 
large magnetic field gradients in the chromosphere. Another avenue is to consider how 
the resolution in the chromosphere and corona limits the width 
of possible spicule and the violence of dynamics at small scales. Other
physical process that might affect the dynamics is the inclusion of a generalized
Ohm's law. A necessary prerequisite in this case is to treat the ionization of hydrogen in 
a time dependent manner, which in and of itself will also have profound effects on 
chromospheric dynamics as discussed in detail by 
\citep[][]{Carlsson:2009ek,Leenaarts:2010fk,Martinez-Sykora:2010vl}.

Nevertheless, we have demonstrated here that the combination of strong Lorentz forces
in particular magnetic field geometries in the upper chromosphere can produce 
high velocity jets of cool material along the magnetic field into the corona that share
several characteristics with observed spicules of type~{\sc ii}. 

\section{Acknowledgments}

This research has been supported by 
a Marie Curie Early Stage
Research Training Fellowship of the European Community's Sixth Framework
Programme under contract number MEST-CT-2005-020395: The USO-SP 
International School for Solar Physics. 
Financial support by the European Commission through the SOLAIRE Network 
(MTRN-CT-2006-035484) and by the Spanish Ministry of Research and Innovation 
through projects AYA2007-66502 and CSD2007-0050 are gratefully acknowledged. 
Supported through grants SMD-07-0434, SMD-08-0743, SMD- 09-1128, SMD-09-1336, 
and SMD-10-1622 from the High End Computing (HEC) division of NASA.

The 3D simulations have been 
run with the Njord and Stallo cluster from the Notur project. We thankfully
acknowledge the computer and supercomputer 
resources by Research Council of Norway through grant 170935/V30 and through 
grants of computing time from the Programme for Supercomputing. 
This work was made possible by NASAÕs High-End
Computing Program. In addition, the simulation presented in this paper
was carried out on the Columbia cluster at the Ames Research
Center. We thank the Advanced Supercomputing Division staff
for their technical support.

To analyze the data we have used IDL and Vapor 
(http://www.vapor.ucar.edu). 

We would like to thank to the referee, whose comments helped us to
improve the manuscript considerably. We also thank Bart de Pontieu for
the nice discussions on the topic of this paper. 

\bibliographystyle{aa}

\end{document}